\documentclass[11pt,a4paper]{article}
\pdfoutput=1
\usepackage{jheppub, hyperref, cleveref}
\usepackage{amsfonts, amsmath, amssymb, latexsym}
\usepackage{graphicx, caption}
\usepackage[mathscr]{eucal} 

\newcommand{\mmax}{m_\mathrm{max}}
\newcommand{\mmin}{m_\mathrm{min}}

\usepackage{comment}

\author[a]{Mauro Pieroni}
\emailAdd{m.pieroni@imperial.ac.uk}
\affiliation[a]{Blackett Laboratory, Imperial College London, South Kensington Campus, London, SW7 2AZ, UK}

\author[b, c]{, Angelo Ricciardone}
\emailAdd{angelo.ricciardone@pd.infn.it}
\affiliation[b]{Dipartimento di Fisica e Astronomia ``G. Galilei'', Universit\`a degli Studi di Padova, via Marzolo 8,
I-35131, Padova, Italy}
\affiliation[c]{INFN, Sezione di Padova, via Marzolo 8, I-35131, Padova, Italy}

\author[d, e]{, Enrico Barausse}
\emailAdd{barausse@sissa.it}
\affiliation[d]{ SISSA, Via Bonomea 265, 34136 Trieste, Italy and INFN Sezione di Trieste}
 \affiliation[e]{IFPU - Institute for Fundamental Physics of the Universe, Via Beirut 2, 34014 Trieste, Italy}

\title{Detectability and  parameter estimation of
stellar origin black hole binaries with next generation gravitational wave detectors}

\abstract{We consider stellar-origin black hole binaries, which are among the main astrophysical sources for next generation gravitational wave (GW) detectors such as the Einstein Telescope (ET) and Cosmic Explorer (CE). Using population models calibrated with the most recent LIGO/Virgo results from O3b run, we show that ET and CE will be capable of detecting tens of thousands of such sources (and virtually all of those present in our past light cone up to $z\lesssim 0.7$ for ET and $z\lesssim 1$ for CE) with a signal-to-noise ratio up to several hundreds, irrespective of the detector design. When it comes to parameter estimation, we use a Fisher-matrix analysis to assess the impact of the  design  on the estimation of the intrinsic
and extrinsic parameters. 
We find that the CE detector, consisting of two distinct $L-$shape interferometers, 
has better sky localization performance compared to ET in its triangular configuration. We also find that the network is typically capable of measuring the chirp mass, symmetric mass ratio and spins of the binary at order of $10^{-5}$,  $10^{-4}$ and $10^{-4}$ fractional error respectively. While the fractional errors for the extrinsic parameters are of order $10^{-2}$ for the sky localization, luminosity distance and inclination.}

\begin{document}

\begin{flushleft}
ET-0036A-22
\end{flushleft}

\maketitle

\section{\sc Introduction}
\label{sec:intro}

The European Strategy Forum on Research Infrastructures (ESFRI) has recently decided to include the Einstein Telescope (ET)
gravitational wave (GW) detector in the 2021 roadmap~\cite{Punturo:2010zz, Hild:2010id}. At the same time the US community is planning to build 
the Cosmic Explorer (CE), a next generation GW detector with better sensitivity than existing LIGO/Virgo interferometers~\cite{LIGOScientific:2016wof}.
Together with LISA~\cite{LISA:2017pwj}, this experimental effort is bound to revolutionize 
our understanding of the gravitational Universe. After the first  
 detection of GWs by the LIGO/Virgo collaboration~\cite{LIGOScientific:2016aoc, LIGOScientific:2018mvr, LIGOScientific:2020ibl, LIGOScientific:2021djp},
 nowadays
the detection rate of binary black holes (BBHs) is around one per week. With next generation detectors, it is expected that  ET at design sensitivity will
perform around $\sim 10^{5}-10^{6}$ BBH detections and $\sim 7\times 10^{4}$ binary neutron star (BNS) detections per year~\cite{Regimbau:2012ir, Belgacem:2019tbw}, i.e. up to several per hour. Moreover, given the one order of magnitude improvement in  sensitivity compared to second generation interferometers, ET or CE will be able to detect BBHs and neutron star - black hole (NSBH) binaries with total mass between 20 and 100 $M_\odot$, up to redshift z $\sim$ 20. For BNS binaries (i.e. total masses $\sim 3  M_{\odot}$), ET or CE should be able to reach $z \sim (2-3)$~\cite{Sathyaprakash:2019nnu, Maggiore:2019uih}.

Such estimates are of course related to the detector configurations/positions and to the astrophysical population. At the time of writing, the final configuration and locations of ET and CE have not been yet decided. For ET, a triangular detector configuration  seems the most plausible option. Concerning the location, at the moment there are two possibilities: one in Sardinia (Italy) and one in the Eusebio region of the Netherlands. On the US side, the shape of the CE interferometers should remain  $\rm{L}$-like, but characterized by longer arm-lengths (in particular, one interferometer of 40 km and one of 20 km).

The increased sensitivity of these detectors will have an impact on the parameter estimation accuracy, both for extrinsic and intrinsic source parameters.
In this paper, we assess the detectability and the parameter estimation accuracy for one of the prime astrophysical sources for ET and CE, i.e, stellar-origin black hole binaries (SOBHBs). We build simulated populations of SOBHBs according to the latest mass functions inferred by LIGO/Virgo~\cite{LIGOScientific:2021psn}, and adopting the IMRPhenomXHM~\cite{Garcia-Quiros:2020qpx} waveform model, which includes higher order modes, and assumes quasi-circular
(i.e. non-eccentric) and non-precessing black hole binaries. Then
we use a Fisher matrix analysis to compare the accuracy in parameter estimation of the two detectors: one case where ET consists of three co-located nested detectors in a triangular configuration placed in Sardinia, and two hypothetical CE detectors consisting of two L-shaped interferometers placed one in Livingston and one in Hanford (where the two LIGO detectors are currently located).

We first estimate the number of observed sources as a function of redshift and  total mass, both for ET and CE separately and for a network consisting of both detectors simultaneously, assuming a threshold in  signal-to-noise ratio (\rm{SNR}) of 20. We also comment on the horizon distance of these detectors, highlighting how the increase in this quantity relative to previous generation detectors will greatly enhance the discovery space also in `exotic' regions of the parameter space (e.g. for putative black holes of sub-solar mass,
 which cannot be produced
by stellar evolution and which could therefore be of primordial origin)~\cite{Bird:2016dcv, DeLuca:2020qqa, Clesse:2016vqa, Domcke:2017fix, Bartolo:2019zvb}.We also look at the distribution of the number of events and at the detected fraction of the astrophysical population, for the two detectors and the network, and we study the SNR distribution of the predicted detections.

We then move to the estimation of the errors on the source parameters, 
both intrinsic and extrinsic. Among the latter, we include the luminosity distance $d_L$, the sky localization $\Omega$, the polarization angle $\psi$ and the inclination angle $\iota$. We quantify the fraction of events that can be detected with a sky location error smaller than 10, 1 and 0.1 square degrees, both at low ($z<2$) and high redshift ($z>2$). Similarly, we calculate the percentage of events that can be detected with a relative statistical error better than $20\%$, $10\%$ and $5\%$ on the luminosity distance. These estimates are relevant
 e.g. to project how GWs can potentially be used for investigating the large-scale structure (LSS) of the Universe~\cite{Laguna:2009re, Jeong:2012nu, Camera:2013xfa, Oguri:2016dgk, Raccanelli:2016cud, Scelfo:2018sny, Scelfo:2020jyw}, for studying the properties of the host galaxies~\cite{Adhikari:2020wpn} and for constraining cosmological parameters~\cite{DelPozzo:2011vcw,Mukherjee:2018ebj, Libanore:2020fim, Libanore:2021jqv}. 
 
We also study the degradation of the luminosity distance estimates due to  weak lensing, as a function of redshift. The latter acts as a systematic bias, which can influence astrophysical and cosmological parameter estimation~\cite{Holz:2004xx,Hirata:2010ba,Bertacca:2017vod,Bertacca:2019fnt}.
 
Regarding the intrinsic source parameters, we derive  projected errors on the  chirp mass $\mathcal{M}_c$, the symmetric mass ratio $\eta$, the two spins $\chi_{1, 2}$, again for ET and CE independently, and for a network of the two.  For a selected event with large SNR in our simulated populations, we also present detailed  Fisher Matrix posterior forecasts. 

The structure of the paper is the following:  in Section~\ref{sec:decandcat} we  specify the assumed ET and CE detector properties, and we explain the procedure adopted for generating our simulated catalogues of sources; in Section~\ref{sec:ex_par} we present our parameter estimation results for the extrinsic  and  intrinsic parameters. A conclusion and an appendix on the adopted  mass function and spin distribution for our simulated populations conclude the paper.

\section{\sc Detector Characterization and Catalogue Generation}
\label{sec:decandcat}
\subsection{Detector properties}
\begin{figure}[t!]
	\centering
	\includegraphics[width=.8\textwidth]{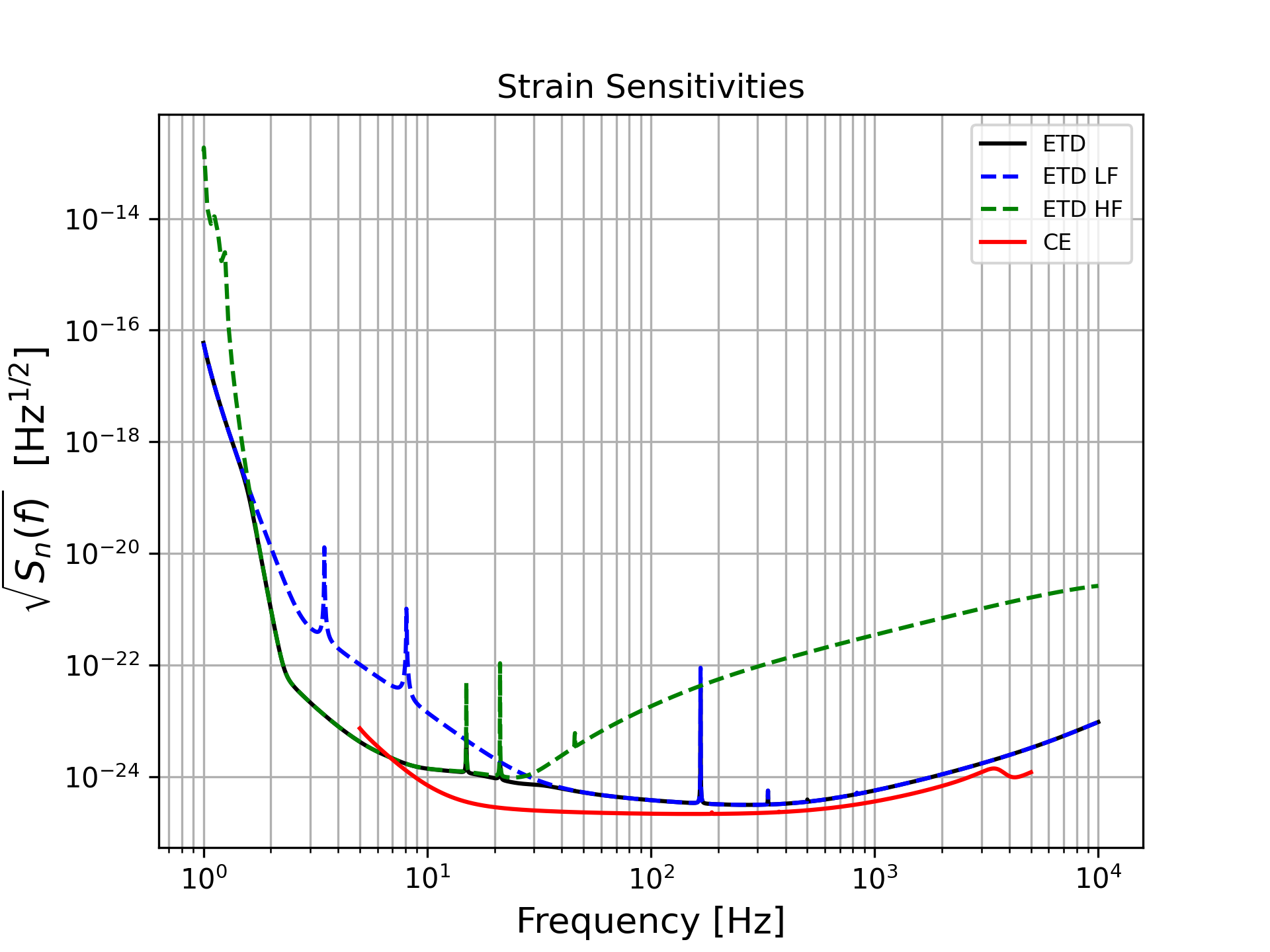}
	\caption{{\it Plot of ET-D and CE strain sensitivities.}} 
	\label{fig:sensitivity}
\end{figure}

ET and CE are next generation interferometers that will have an order of magnitude better sensitivity and a wider accessible frequency band  (from 3 $\rm {Hz}$ to many $\rm {kHz}$) than current GW detectors. The final configuration and the location of the detectors have not been  finalized yet. However, in this paper, considering the most recent public specifications, we consider the most plausible case where ET consists of three co-located nested detectors in a triangular configuration placed in Sardinia\footnote{Each individual detector will be composed of two interferometers, one specialised for detecting low-frequency GWs and one for high-frequency GWs, forming a so-called xylophone configuration. The two interferometers will be of Michelson type with an opening angle of 60 degrees. Since the two detectors have a similar geometry, they will share common tunnels.
As for the exact location, we consider the Sos Enattos mine in the city of Lula in Sardinia (N $ 40^{\circ}\, 26'$, E $9^{\circ}\, 26'$).},
while for CE we consider the case of two L-shaped interferometers placed one in Livingston and one in Hanford,  where the two LIGO detectors are currently located~\footnote{Livingston: N $30^{\circ} \,33'$, W $90^{\circ}  \,46'$, Hanford: N $46^{\circ} \,27'$, W $119^{\circ}  \,24'$.}. In the ET case, we consider a 10 km arm length, while for CE we consider 40 and 20 km respectively. We focus on the ET-D noise power spectral density~\cite{Punturo:2010zz, Hild:2010id}, while for CE we consider that of \cite{LIGOScientific:2016wof, ce}. In Fig.~\ref{fig:sensitivity}, we show the ET and CE strain sensitivities taken from~\cite{ET_sens}. For future reference, we also show the power law sensitivities \cite{Thrane:2013oya, Caprini:2019pxz} for stochastic backgrounds, computed considering one year of mission and an SNR threshold of 10 for the background.
 In this study we do not consider networks that include current generation interferometers nor their upgraded $\rm{A}+$  versions, but we just focus on next generation ones. 
 The $\rm{A}+$ detector generation will have an improvement in sensitivity of a factor $\sim$ 2 to 4 depending on the frequency, compared to current generation detectors, driving the transition to next generation detectors such as ET and CE.  For a detailed analysis including networks consisting also of $\rm{A}+$ detectors, see \cite{Borhanian:2022czq}.

 \begin{figure}[t!]
	\centering
	\includegraphics[width=.8\textwidth]{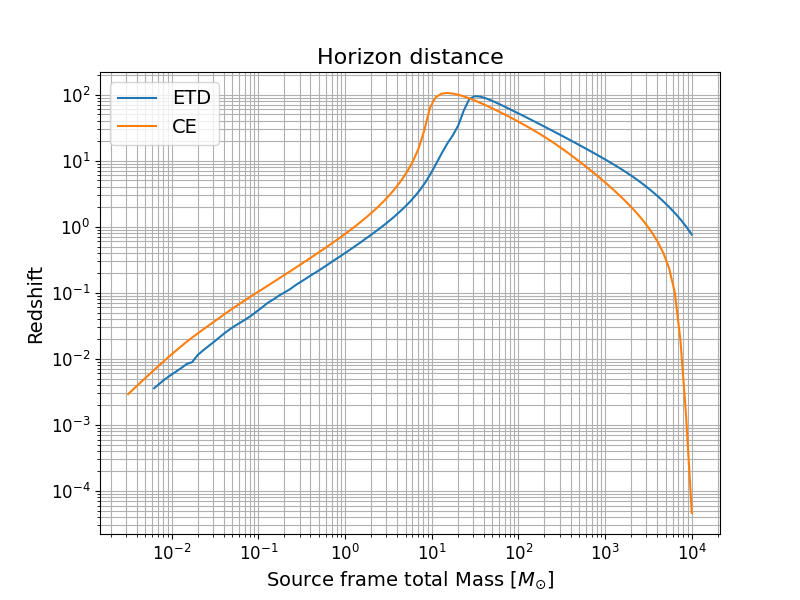}
	\caption{\it Horizon distance plot for ET and CE for equal mass system of black holes and assuming that all the events above $\rm{SNR}_{th} = 20$ are detectable.}
	\label{fig:horizon_distance}
\end{figure}

\subsection{Waveforms, astrophysical black hole populations and catalogue generation}
\label{sec:catfish}
In our analysis we adopt the IMRPhenomXHM~\cite{Garcia-Quiros:2020qpx}  waveform model, as implemented in PyCBC~\cite{alex_nitz_2021_5347736}. 
This waveform class includes higher order modes, and assumes
quasi-circular (i.e. non-eccentric) and
non-precessing black hole binaries, with $\chi_{1,2}$ the dimensionless spin 
parameters projected on the orbital angular momentum axis.

Overall, we therefore
characterize our simulated black hole binaries by the following parameters:
\begin{equation}
   z,\, m_{1},\, m_{2}, \,  \chi_{1}, \, \chi_{2}, \, \theta,\, \phi,\,\psi,\, \iota,\, \tau_{c},\, \phi_0\,,
\end{equation}
where $z$ is the redshift of the event, $\theta$ and $\phi$ are respectively the declination and right ascension of the source, $\psi$ is the polarization angle, $\iota$ is the inclination angle, $\tau_c$ is the coalescence time and $\phi_0$ is the initial phase of the binary. Below, we will perform a Fisher matrix analysis on these 11 parameters.\\

For a given detector, whose instrumental response can be characterized through the sky-dependent pattern functions $F^{+/\times}_{ij}$ (for the definition see for example~\cite{Maggiore:1999vm}), the signal can be expressed as:
\begin{equation}
    h = F^{+}_{ij} h^+_{ij} + F^{\times}_{ij} h^\times _{ij} \; .
\end{equation}
The (optimal) SNR for a given source can then be computed as~\cite{Cutler:1994ys} \emph{i.e.} 
\begin{equation}
    \textrm{SNR}^2 =  \left( \left. h \right| h \right)  \;, 
\end{equation}
where $h$ is the waveform for the event, and $\left( a | b \right)$ denotes the noise weighted inner product:
  \begin{equation}
  \label{eq:noise_weighted}
    (a|b) = 2 \int_{f_1}^{f_2} \frac{a(f)b^*(f)+a^*(f)b(f)}{S_n(f)} \; \textrm{d} f \, ,
\end{equation}
 where $S_n(f)$ is the detector strain sensitivity and $f_1$, $f_2$ are chosen to be the minimum and maximum frequencies of the detector's range.\\

\begin{figure}[t!]
\centering
	\includegraphics[width=.8\textwidth]{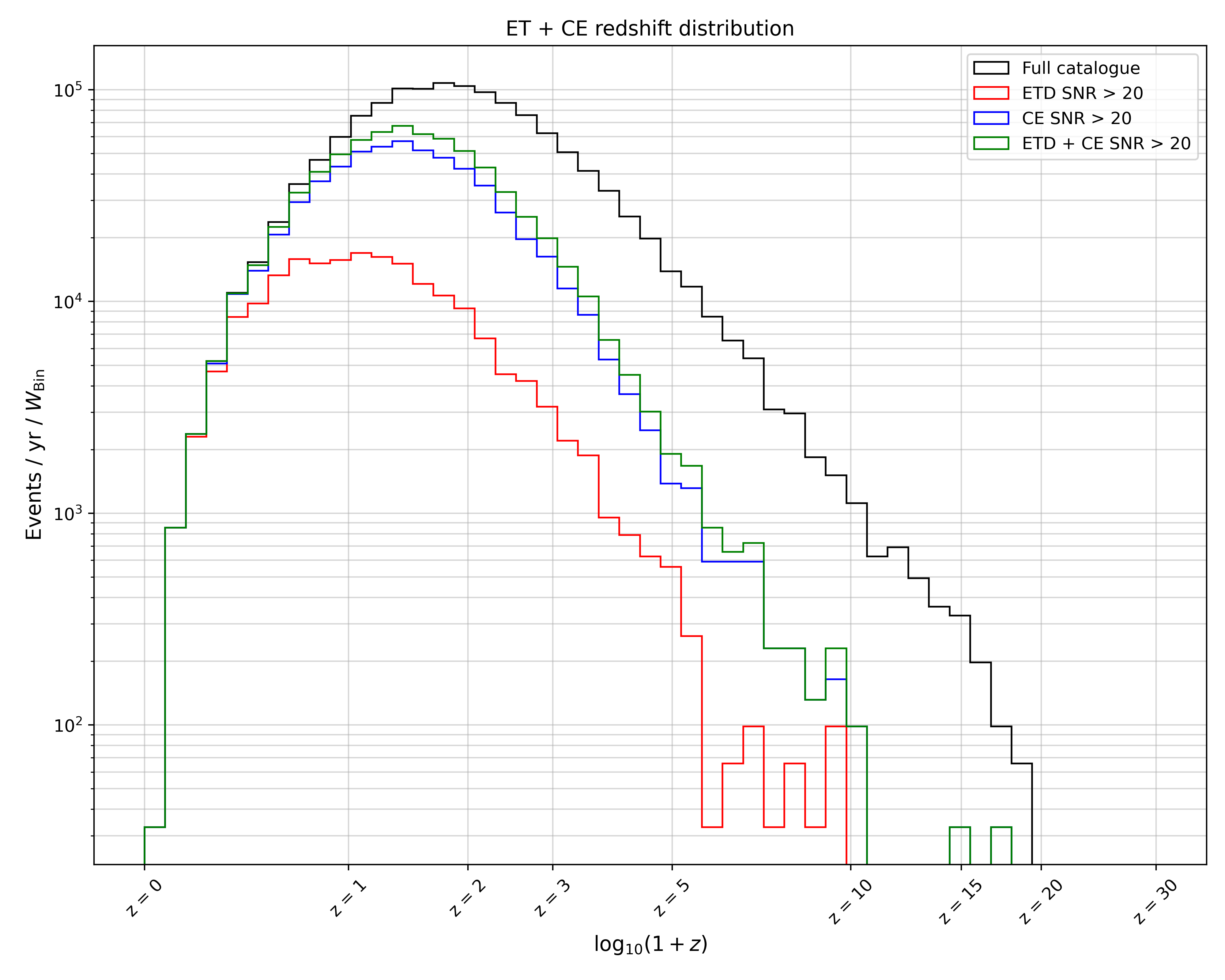}
	\caption{\it Total number of detected black hole mergers as a function of redshift, for ET, CE and the detector network ET+CE. Also shown is the intrinsic number of events that would be detected with an infinitely sensitive detector (``full catalogue'').}
	\label{fig:Fraction}
\end{figure} 

\begin{figure}[t!]
\centering
	\includegraphics[width=.8\textwidth]{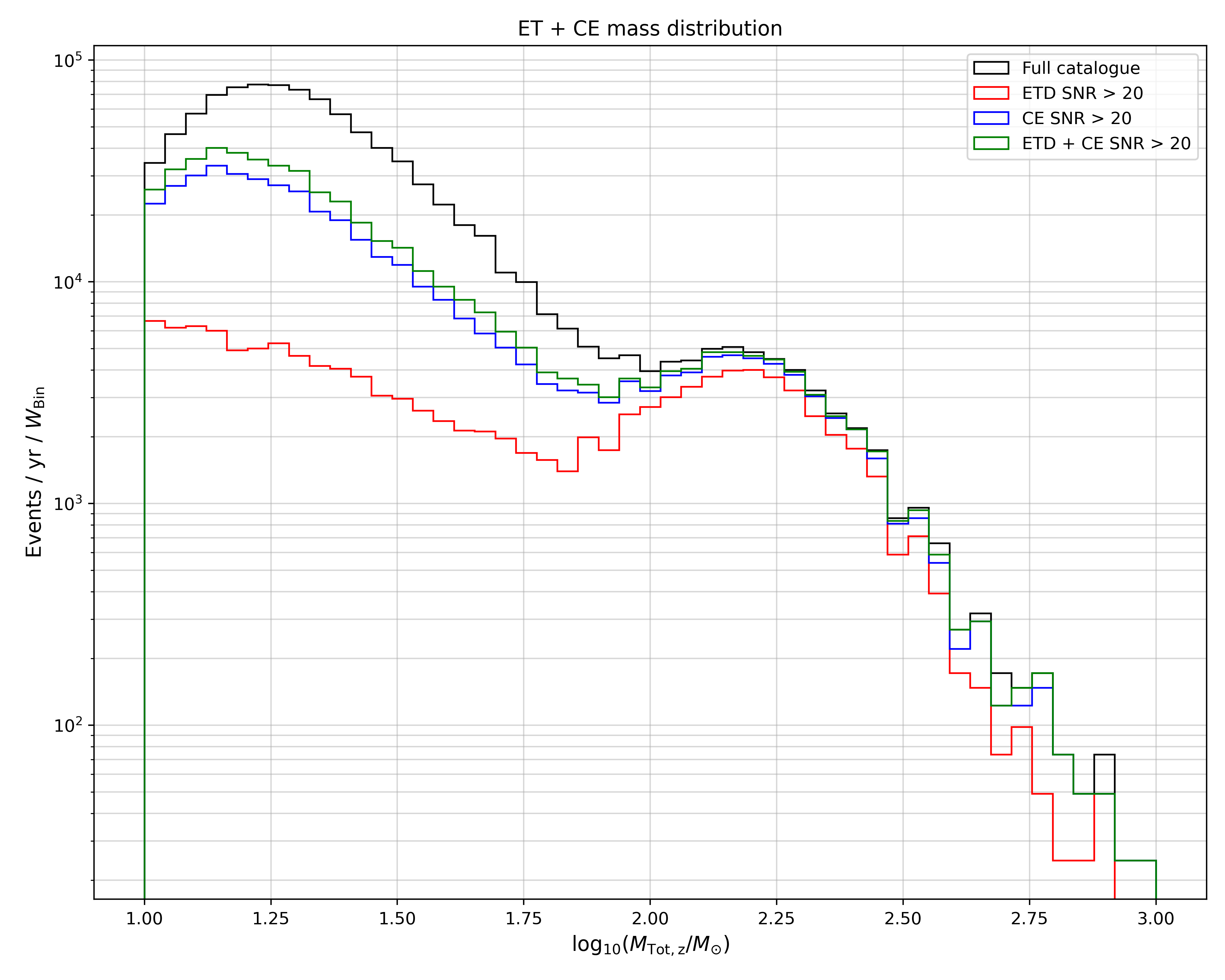}
	\caption{\it  Total number of detected black hole mergers as a function the detector frame total mass, for ET, CE and the detector network ET+CE.}
	\label{fig:Frac_mass}
\end{figure}

\begin{figure}[t!]
	\begin{center} \includegraphics[width=.6\textwidth]{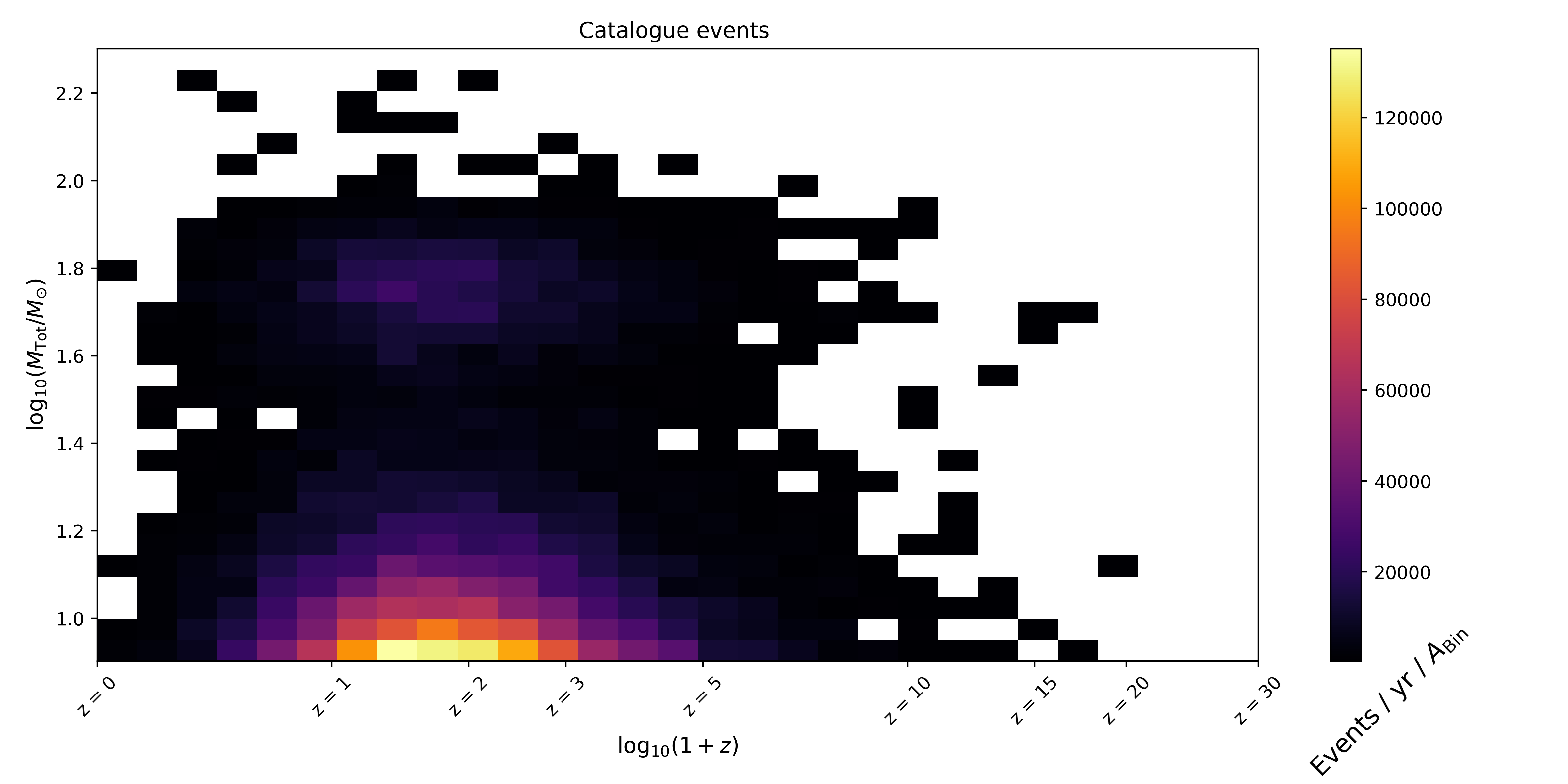}
	\end{center}
	\includegraphics[width=\textwidth]{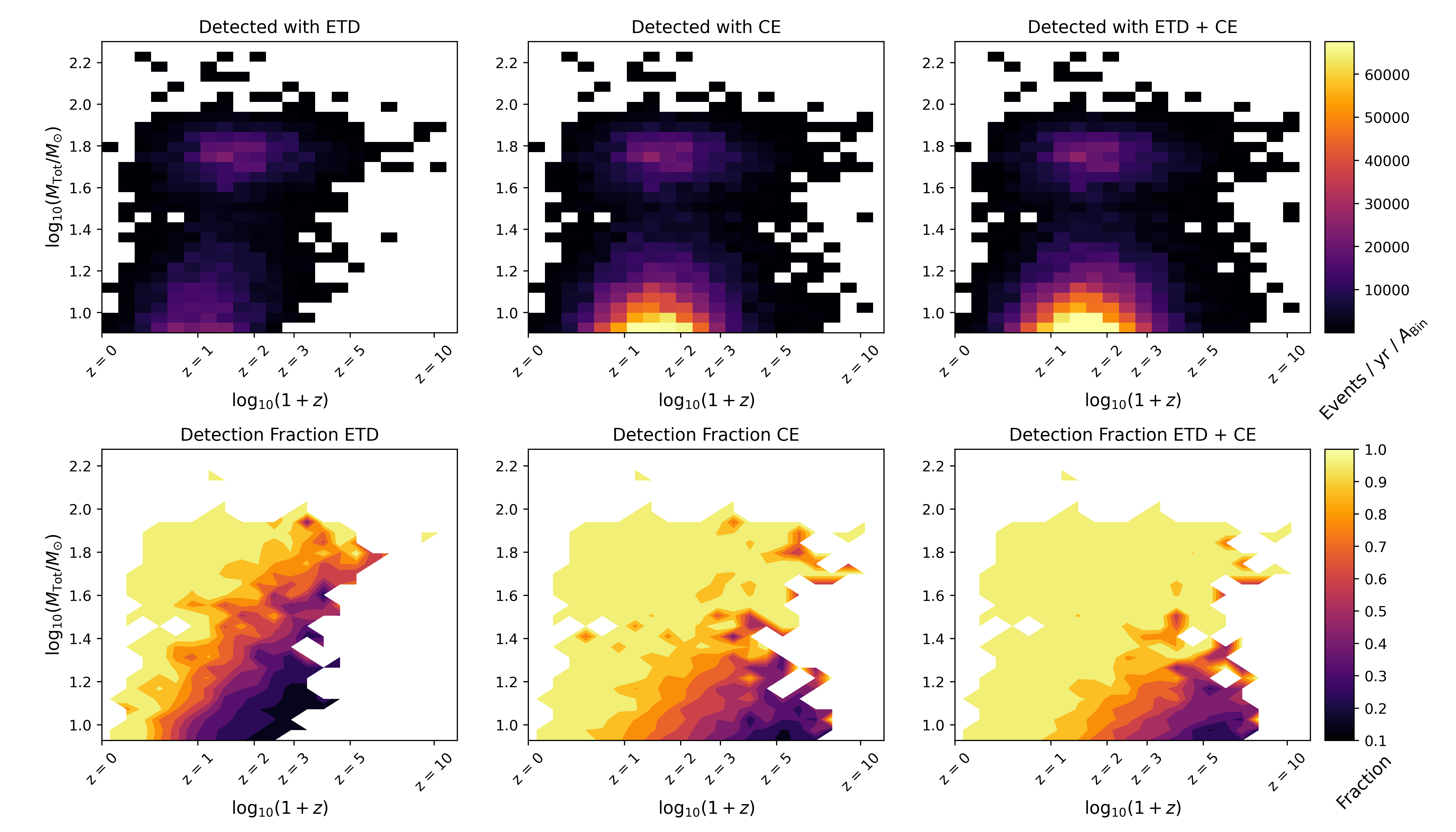}
	\caption{\label{fig:sensitivities}{\it Fraction of the astrophysical population of black hole binaries that is detected, with the ET, CE or the network ET+CE.}}
\end{figure}
\begin{figure}[ht!]
\centering
	\includegraphics[width=.8\textwidth]{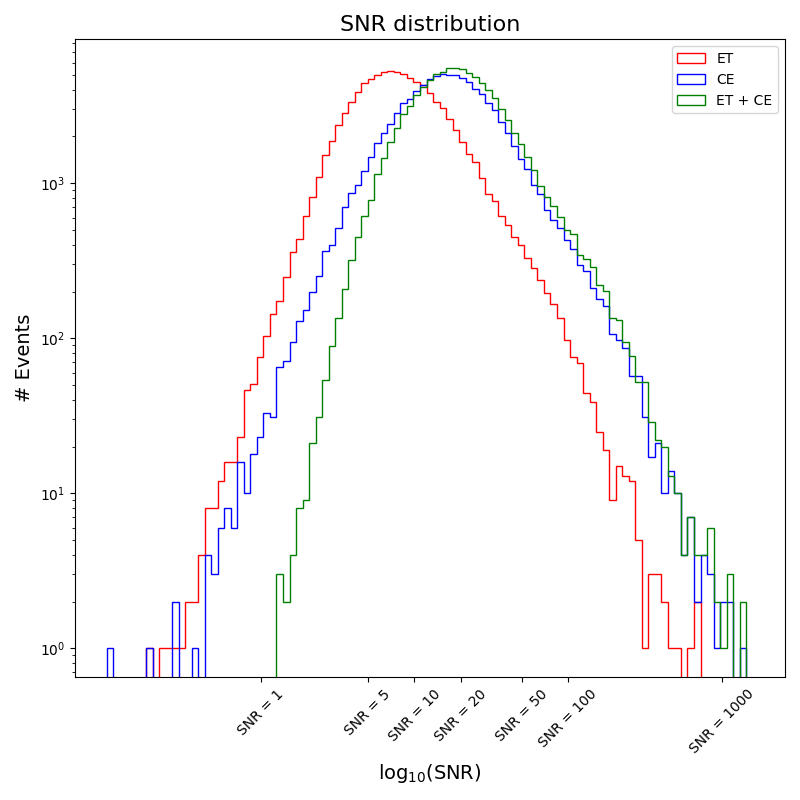}
	\caption{\label{fig:SNR_mass_z} SNR distribution for  ET, CE, and the network ET+CE.}
\end{figure}
\begin{figure}[ht!]
	\centering
	\includegraphics[width=.8\textwidth]{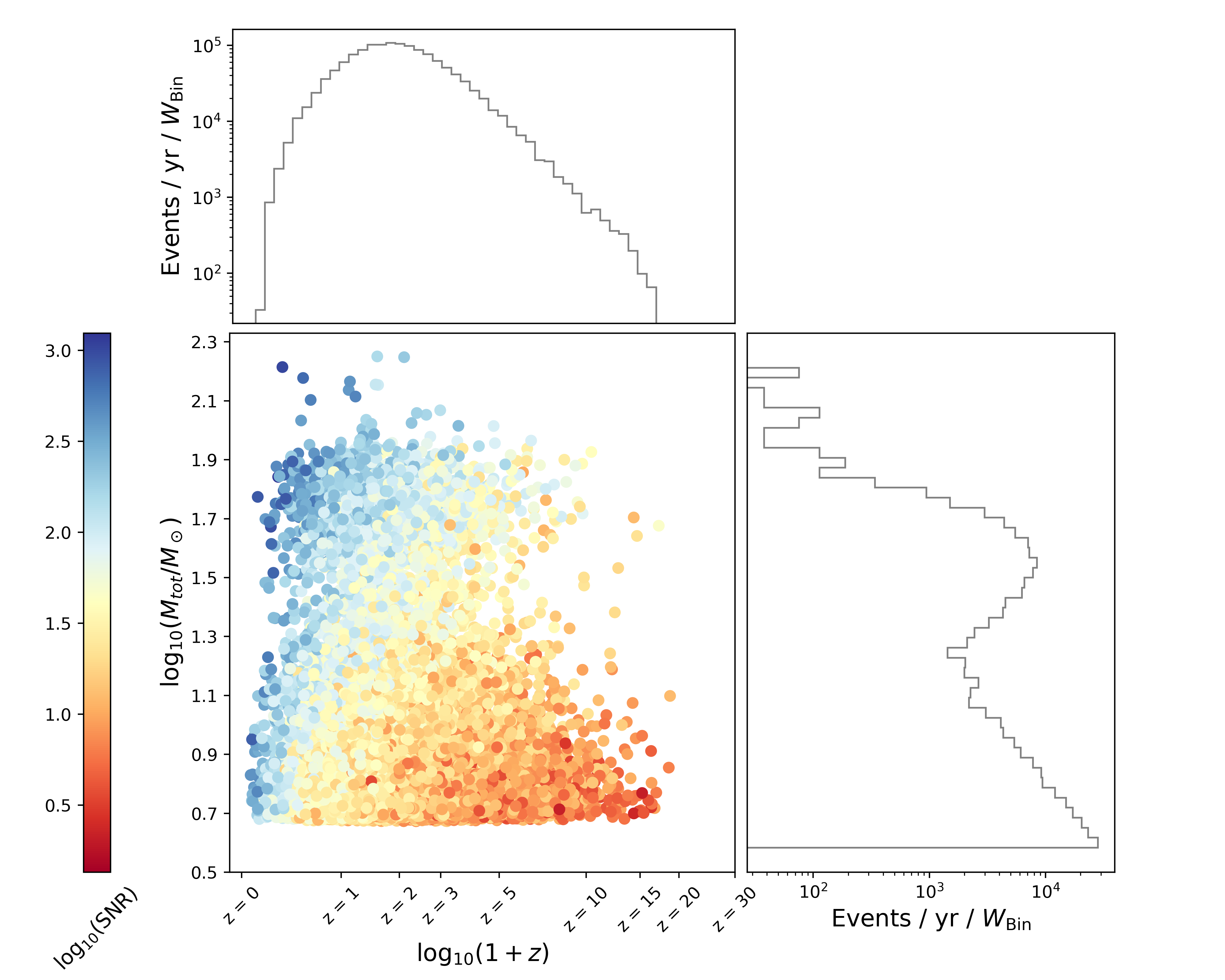}
	\caption{{\it The distribution 
			of source-frame mass and redshift for 
			one universe realization, together with the SNR (color coded), for the network ET+CE.}}
	\label{fig:SNR_mass_z_2} 
\end{figure}

We simulate catalogues of SOBHB systems by considering the latest population models from LIGO/Virgo $\rm{O}3b$ run~\cite{LIGOScientific:2021psn} and the latest cosmological parameters constraints from~\cite{Planck:2018vyg}. 
In more detail, we consider a ``power-law plus peak'' mass function as described in appendix~\ref{app:astr_mod}, with a mass range $m_1, m_2 \in [2.3, 100.0] \,M_\odot$. For the spin distribution, we use the LIGO/Virgo {\it default} spin model that we summarize in appendix~\ref{app:astr_mod}. For the SOBHB merger rate, since ET and CE will reach large redshift, we have assumed that the merger rate tracks the Madau \& Dickinson star formation rate as a function of redshift~\cite{Madau:2014bja} convolving this with the time delay between the formation of the binary and its merger. For the time delay distribution we have considered an inverse power-law~\cite{Mapelli:2017hqk,Borhanian:2022czq}
\begin{equation}
p(t_d) = \frac{1}{  \ln\left( t_d^{\rm max} / t_d^{\rm min} \right) t_d}\, ,
\end{equation} 
between $t_d^{\rm min}= 10\, \rm Myr$ and $t_d^{\rm max}= 10\, \rm Gyr$. 
This means that the merger rate increases up to $z\sim2$  and then decreases again at higher redshift. This dependence has  also been considered in the LIGO/Virgo papers on the GW stochastic background~\cite{KAGRA:2021kbb}. We considered just one type of  population of astrophysical objects (see~\cite{Singh:2021zah} for an analysis which take into account different populations). The normalization of the merger rate is chosen such that its local value at $z =0$ matches with the latest LIGO and Virgo observations (i.e., $R_{0}=17.3$ $\rm{Gpc}^{-3} \rm{yr}^{-1}$,  ${\it{k}}$ = $2.9$)~\cite{LIGOScientific:2021psn}.   We focus on the redshift interval $0<z<15$, but our results are insensitive to the maximum redshift that we consider. For simplicity, we do not consider the possibility of two or more overlapping signals and that we are able to identify all the events~\footnote{Techniques such as those employed in the LISA Data Challenge~\cite{Caprini:2019pxz, Flauger:2020qyi, Pieroni:2020rob, Boileau:2021sni} may be beneficial to treat that case, or the inclusion of anisotropies could help~\cite{Bellomo:2021mer}}. 

In terms of redshift reach, next generation detectors will
be significantly better than LIGO/Virgo: 
comparing to the latter, which are sensitive to
events up at redshift $z \lesssim 1$, ET and CE should reach up to $z \simeq 20$, depending on the mass~\cite{Maggiore:2019uih}, potentially probing the dark era of the Universe preceding the birth of the first stars, and opening the possibility to test a possible primordial origin for, at least part of, the BH population. In particular, in Fig. \ref{fig:horizon_distance} we show the detector reach for equal mass, sky and inclination averaged binaries, in terms of the cosmological redshift, as a function of the source-frame total mass of the coalescing BHB
and considering our default SNR threshold of 20.
As can be seen, ET and CE can reach a similar redshift, however ET will be able to detect BHs with larger masses (up to $\sim 10^3 M_\odot$ up to $z\sim10$). On the other hand, very small source-frame total mass sources (below $1M_\odot$), which cannot be produced by stellar evolution and which
could therefore point at the existence of primordial black holes,
can be observed up to $z\approx .5$ (1) for ET (CE), if they exist.

Note that the large number of detected events by ET or CE, which we will show below to be of order $10^4 $ SOBHBs at high redshift and with a quite high SNR, will allow for
gaining insight on stellar evolution and star formation (e.g. 
the impact of the metallicity)~\cite{Santoliquido:2020axb}. Besides this, the large statistics will also allow to use GW sources in combination with other cosmological probes like Large-Scale Structure (LSS), to shed light on formation scenarios \cite{Raccanelli:2016cud, Scelfo:2018sny}, clustering properties \cite{Calore:2020bpd, Libanore:2020fim, Scelfo:2020jyw}, and test of General Relativity/Modified Gravity \cite{Bertacca:2017vod,Belgacem:2018lbp,Mukherjee:2019wcg}.

With these assumptions, we generate 
mock catalogues of SOBHB mergers for an observation time of one year.
In Fig. \ref{fig:Fraction} we show the number of generated events (``full catalogue'') and the ones detected with SNR $> 20$ for the two detectors separately and for a network ET+CE, as a function of redshift. In the case of the network we define an event {\it detectable} if it is detected with {\rm SNR} larger than 20 in both detectors. We can note  that at low redshift the difference between the two detectors is rather small, while starting from $z\sim 1$ CE is able to detect more events. However, when the network ET+CE is considered,  essentially all the events up to $z\sim 1.5$ are detectable. For comparison, at the present LIGO/Virgo sensitivity, the number of detected BBH events with {\rm SNR} $>20$ is around four~\cite{LIGOScientific:2021psn}.

We then show in Fig. \ref{fig:Frac_mass} the distribution of events per year as a function of the detector-frame total mass. Note that the distribution of events shows a double peak structure, which reflects the power-law+peak mass intrinsic mass distribution adopted in this analysis. We can see that both ET and CE can detect a large number of GW sources with large masses. Some of these may also be observable (during their inspiral phase) in LISA~\cite{Sesana:2016ljz}, which would greatly enhance the 
possibility to use these sources to test General Relativity~\cite{Barausse:2016eii,Toubiana:2020vtf} or to
detect interactions with the surrounding astrophysical environment~\cite{Caputo:2020irr,Toubiana:2020drf}~\footnote{Even if undetected 
as resolved events, these SOBHBs may also appear in LISA as a stochastic background~\cite{Pieroni:2020rob,Karnesis:2021tsh, marcoccia2022}}.

In order to assess the accuracy with which the two detectors can estimate SOBHB parameters, we use a Fisher matrix analysis~\cite{Cutler:1994ys,Vallisneri:2007ev, Cho:2014iaa,Nitz:2021pbr}. The latter is
computationally efficient and can be run on thousand of sources, while 
being accurate only 
in the limit of high SNR~\cite{Cutler:1994ys}.  The Fisher matrix $\Gamma$ for the signal $h$ has the following  elements~\cite{Cutler:1994ys}
\begin{equation}
    \Gamma_{ij}=\left.\left(\frac{\partial h}{\partial \theta^{i}}\right|\frac{\partial h}{\partial \theta^{j}}\right)\,,
\end{equation}
where we are using the noise weighted inner product of~\cref{eq:noise_weighted} and $\theta_{i}$ is the vector of source parameters. The derivatives with respect  to all the parameters are computed analytically, except for the two masses $m_1$, $m_2$ and the spin parameters for which we use a  fourth order finite difference scheme with Richardson extrapolation to vanishing step. Tests of convergence and robustness are performed on the fly trough our code.
 From the Fisher matrix we can build the covariance matrix $\Sigma$ by taking its  inverse,  $\Sigma = \Gamma^{-1}$. In order to stabilize this numerical operation, we condition the parameters which are likely to have the lowest information \emph{i.e.} the initial phase $\phi_0$, the sky localization parameters $\theta$ and $\phi$, the inclination $\iota$ and the phase $\psi$.
 In practice, this corresponds to adding a small number, which we set to be $\epsilon=10^{-10}$, on the diagonal entries of the FIM corresponding to possibly poorly constrained parameters. This is equivalent to assuming
 a loose Gaussian prior with variance $1/\epsilon$ on those parameters. We have also tested that our choice of prior does not affect the parameter reconstruction.
 
 The estimated statistical error on a parameter $\Delta \theta_{i}$ is then computed by extracting the corresponding diagonal element of the covariance matrix, while the error on combinations of the parameters is
  computed by performing standard  error propagation.  \\
In our analysis we focus in particular on the error on the following parameters
\begin{itemize}
    \item error on the chirp mass $\Delta {\cal M}_{c,z}$, where ${\cal M}_{c,z}=(m_1 m_2)^{3/5}/(m_1+m_2)^{1/5}$;
    \item error on the symmetric mass ratio $\Delta \eta$, where $\eta=(m_1 m_2)/(m_1+m_2)^{2}$;
    \item error on the sky location, related to the errors on the $\theta$ and $\phi$ angles by the relation~\cite{Barack:2003fp, Cutler:1994ys, Zhao:2017cbb, Yu:2020vyy}
     $$\Delta \Omega= 2 \pi \sin \theta \sqrt{(\Delta \theta \Delta \phi)^2 - (\Sigma^{\theta \phi})^{2}}\,;$$
    \item  error on the luminosity distance $\Delta d_{L}$.
     \item  error on the two dimensionless spins $\Delta \chi_1$ and $\Delta\chi_2 $;
      \item  error on the inclination angle $\Delta\iota$.
\end{itemize}
In what follows, we will also compare the statistical error on $d_L$
with the systematic error arising from  weak lensing, which we compute by using the fitting formula of \cite{Hirata:2010ba} (used also in~\cite{Tamanini:2016zlh} for LISA analyses):
    \begin{equation}
     \sigma_{lens}(z)= d_L (z)\times  0.066 \left(\frac{1-(1+z)^{-0.25}}{0.25}\right)^{1.8}\,.
     \label{eq: wl}
     \end{equation}

\begin{figure}[ht!]
	\centering
	\includegraphics[width=.85\textwidth]{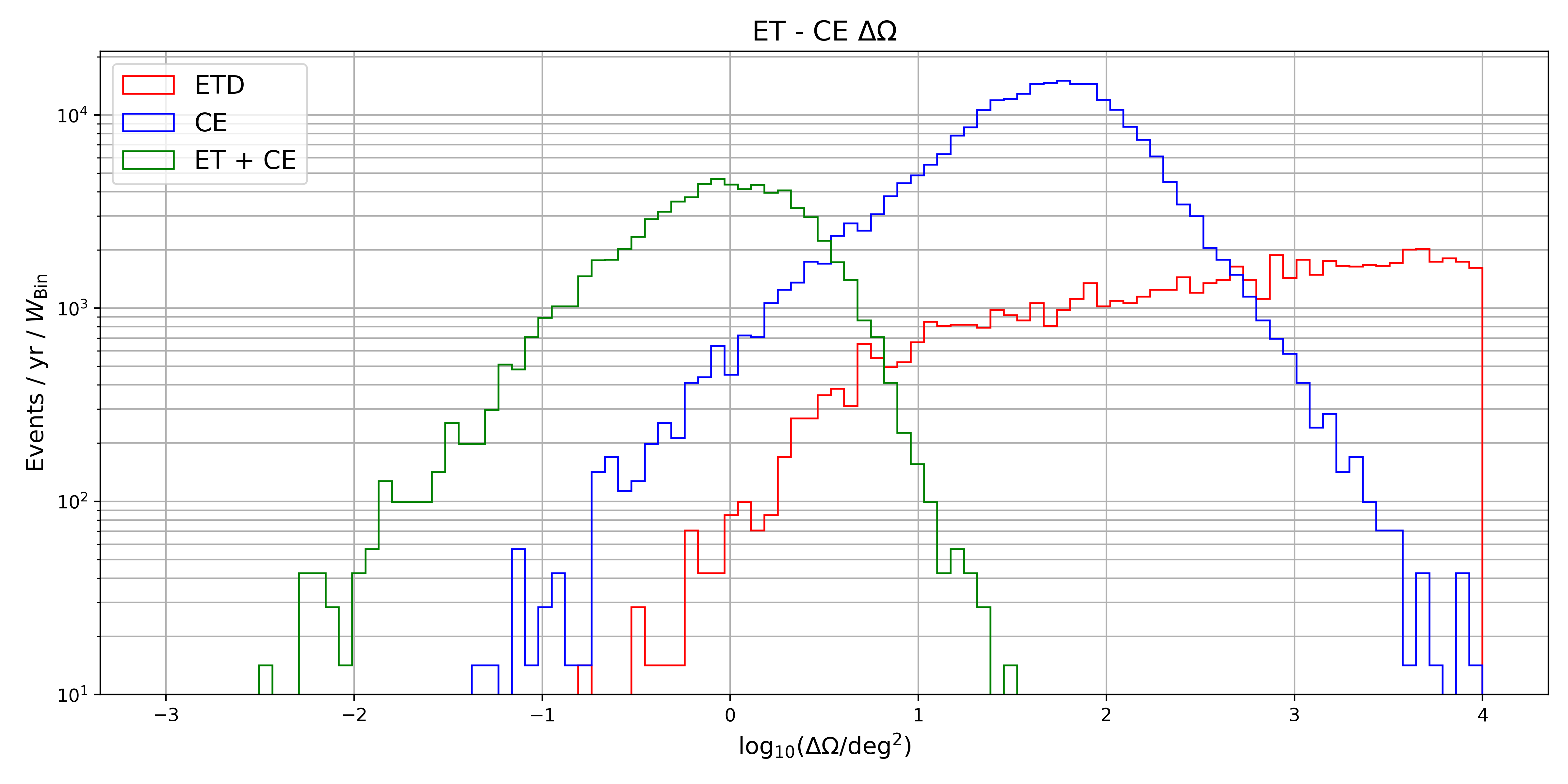}
	\includegraphics[width=.85\textwidth]{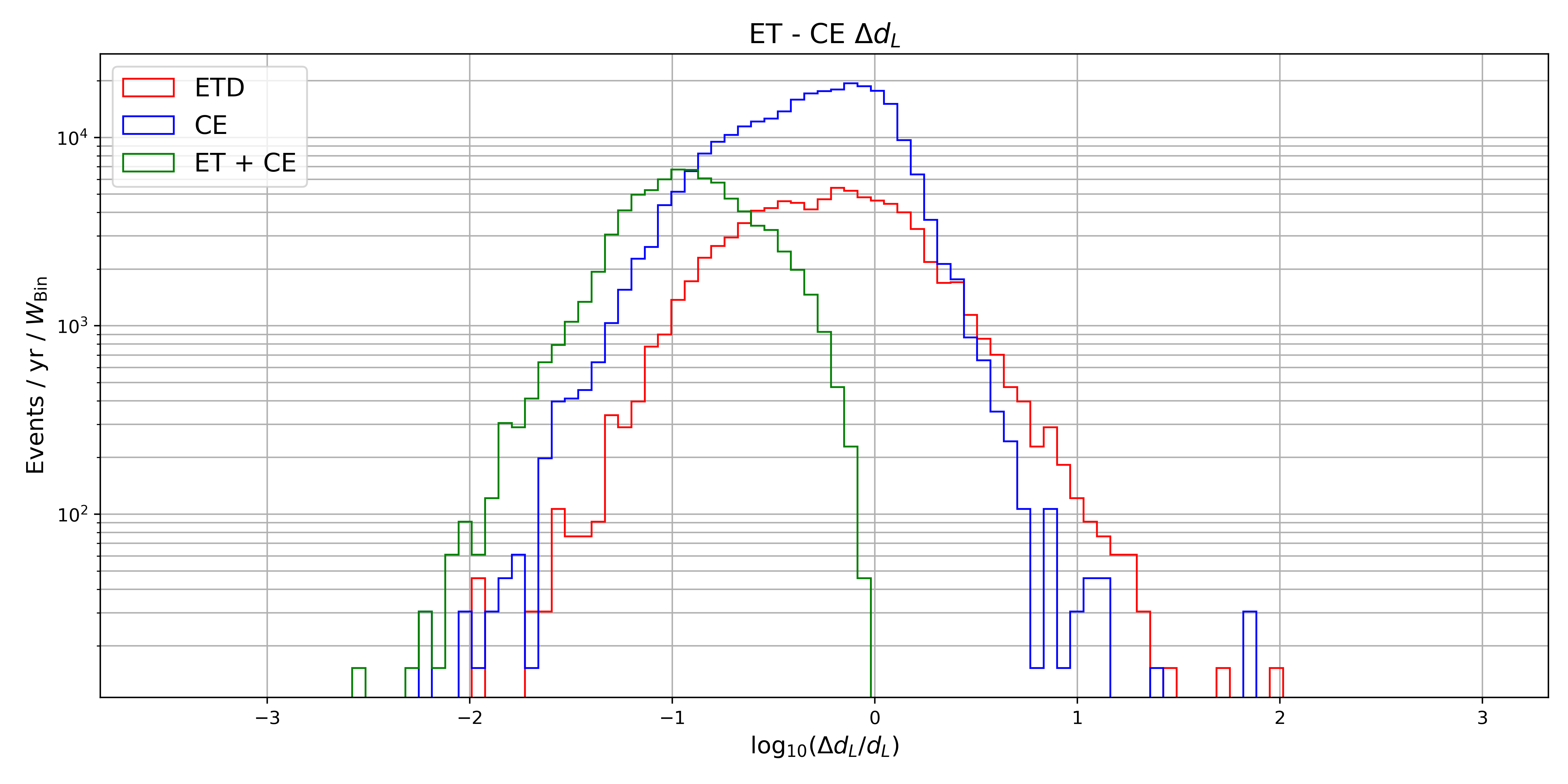}
	\includegraphics[width=.85\textwidth]{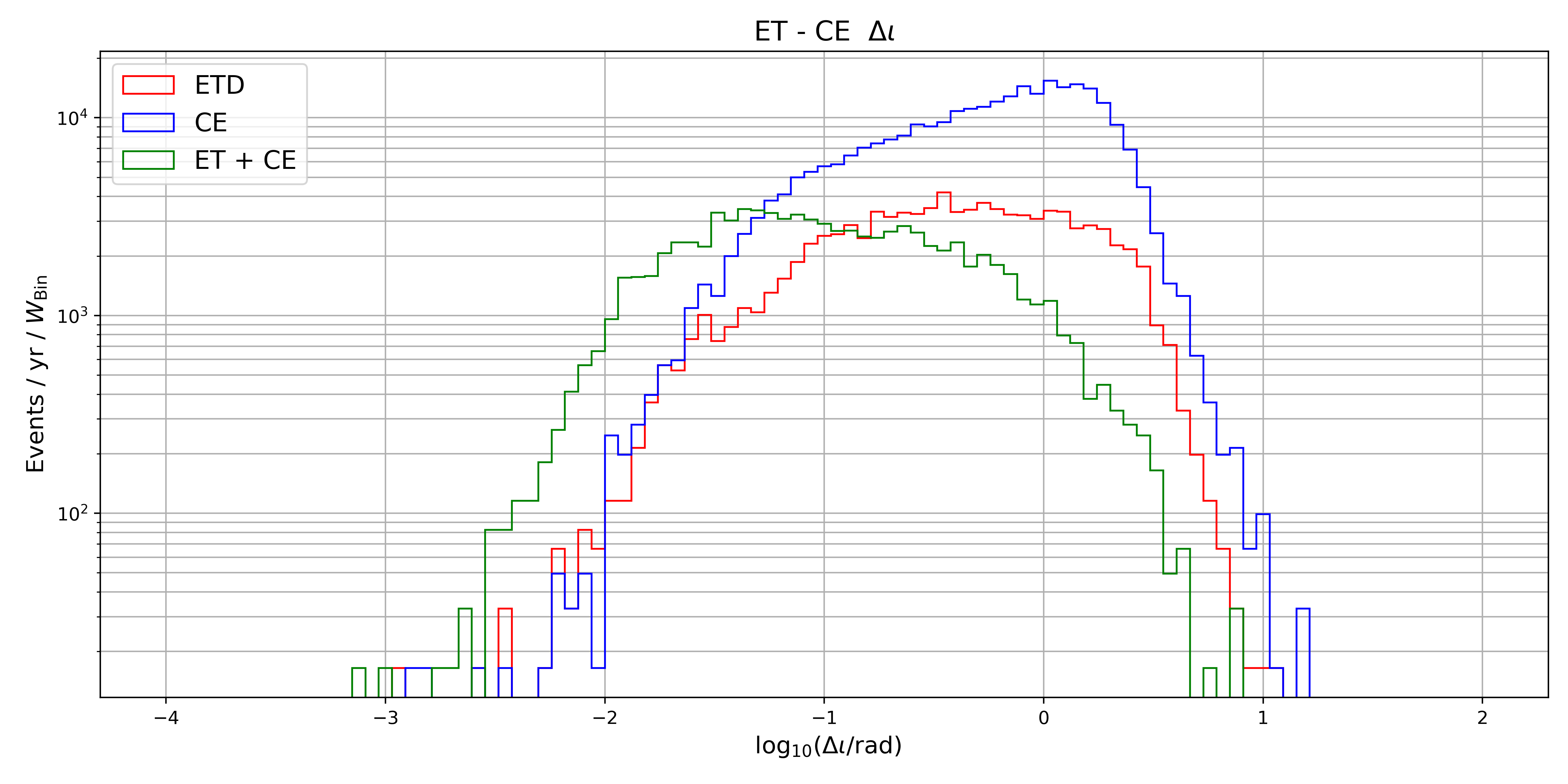}
	\caption{{\it Projected errors on the measurements of the sky position, luminosity distance and inclination, for the detected events (SNR $>20$) in one year for ET, CE and a network ET+CE.}}
	\label{fig:extrinsic}
\end{figure}

\begin{figure}[ht!]
	\centering
	\includegraphics[width=.85\textwidth]{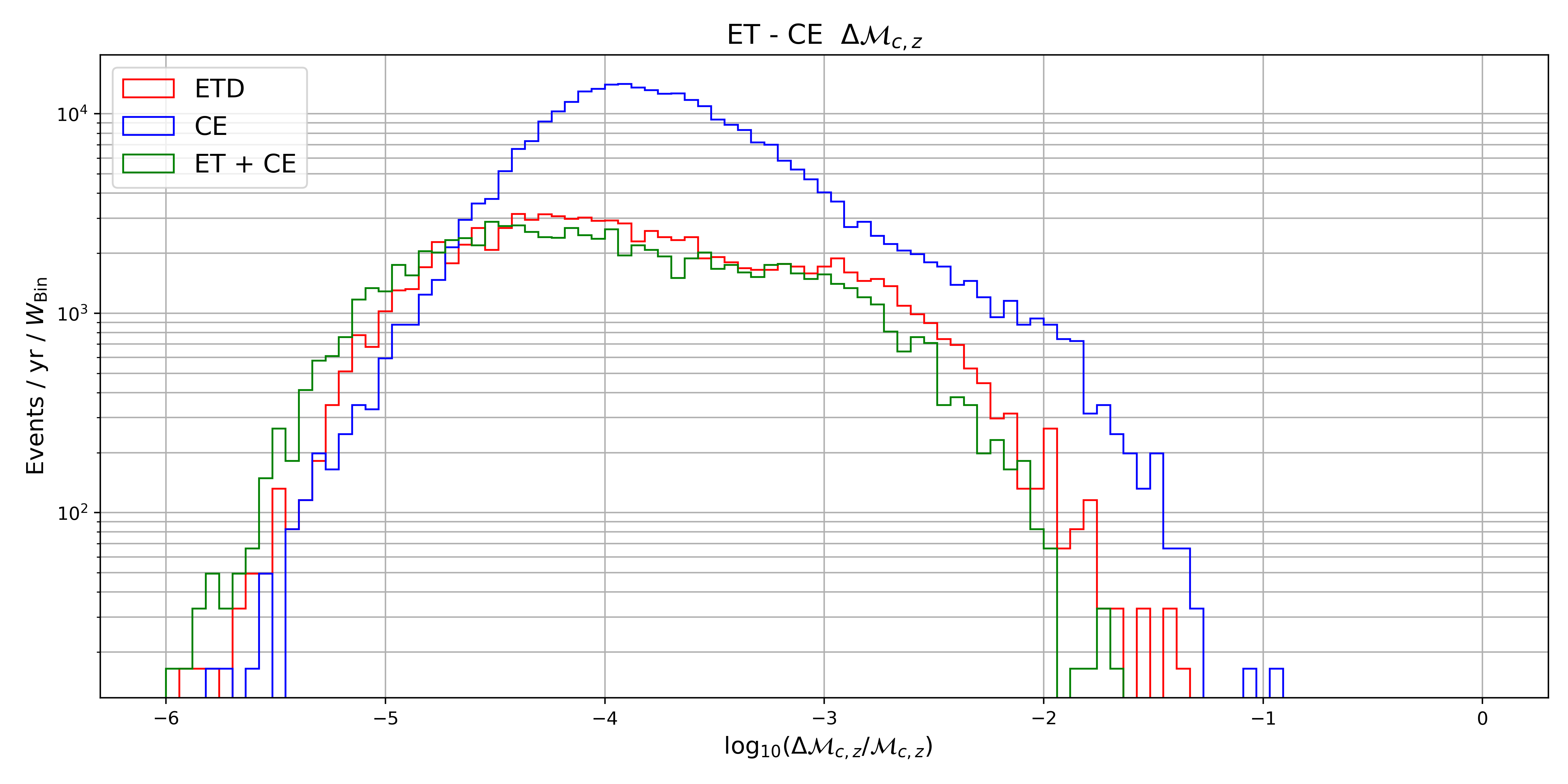}
	\includegraphics[width=.85\textwidth]{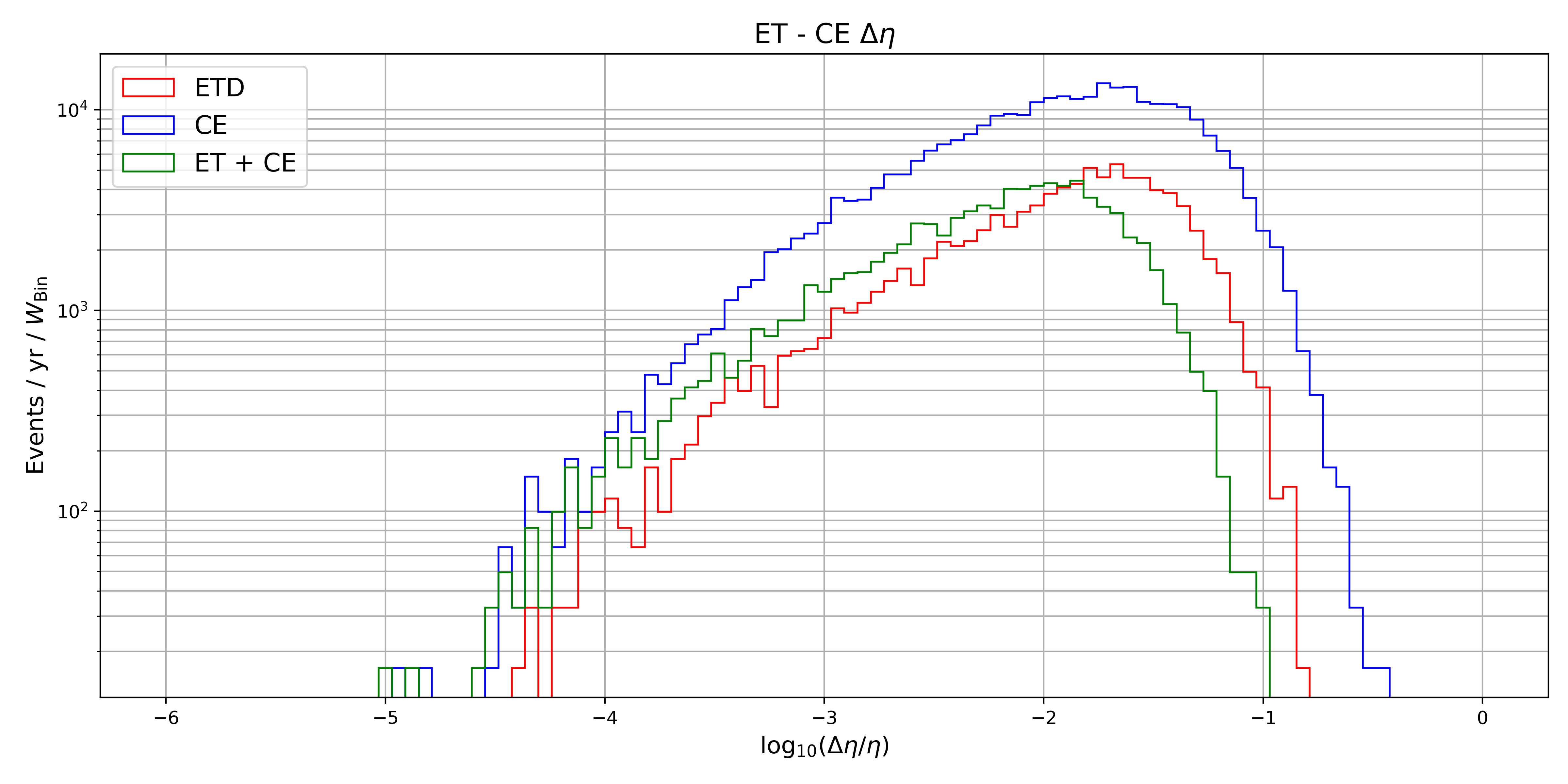}
	\includegraphics[width=.85\textwidth]{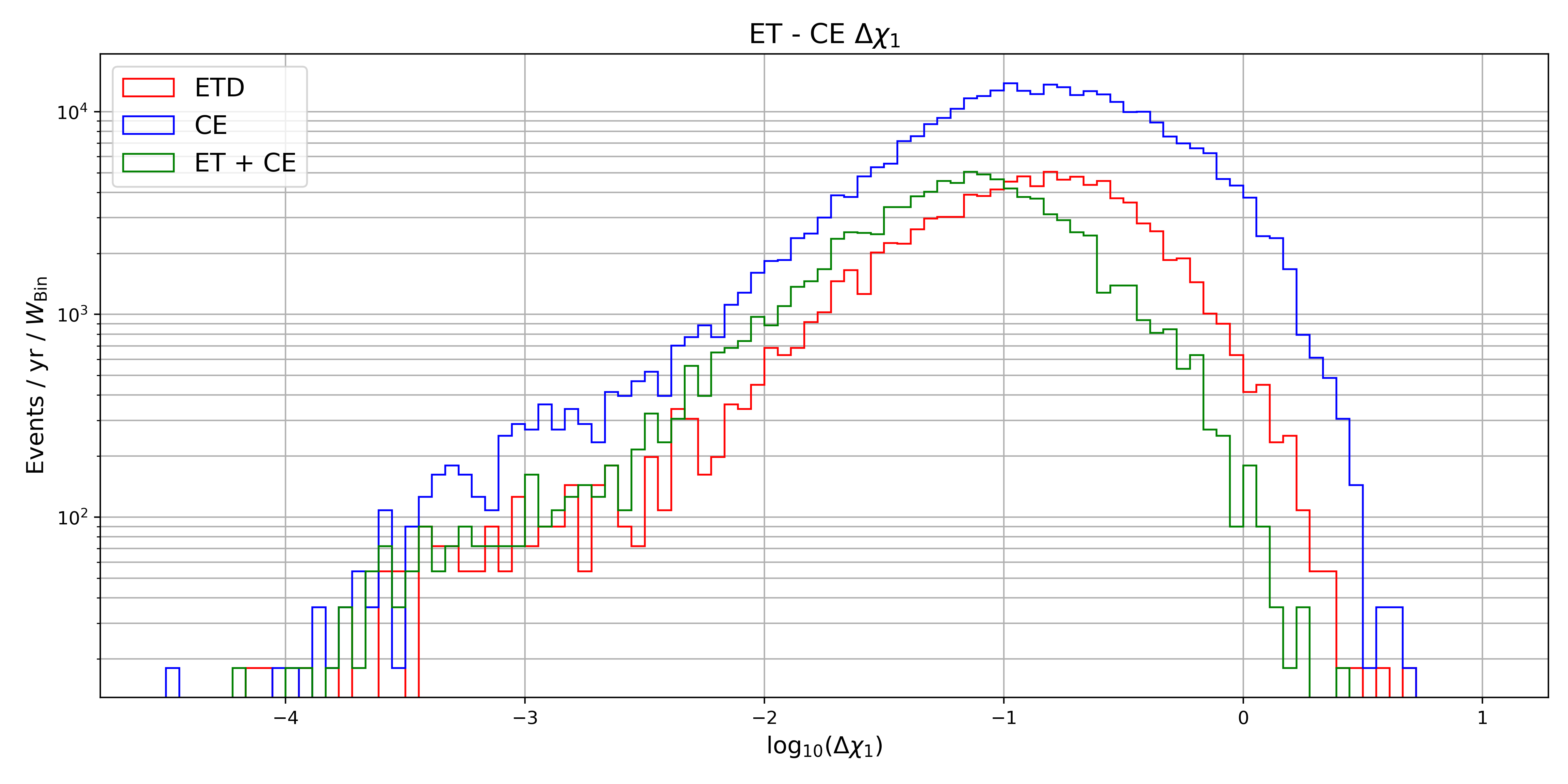}
	\caption{\label{fig:intrinsic}{\it Projected errors on the measurements of the intrinsic parameters, i.e. the
			(detector-frame) chirp mass, symmetric mass ratio and spins.}}
\end{figure}

\begin{figure}[ht!]
	\centering
	\includegraphics[width=.85\textwidth]{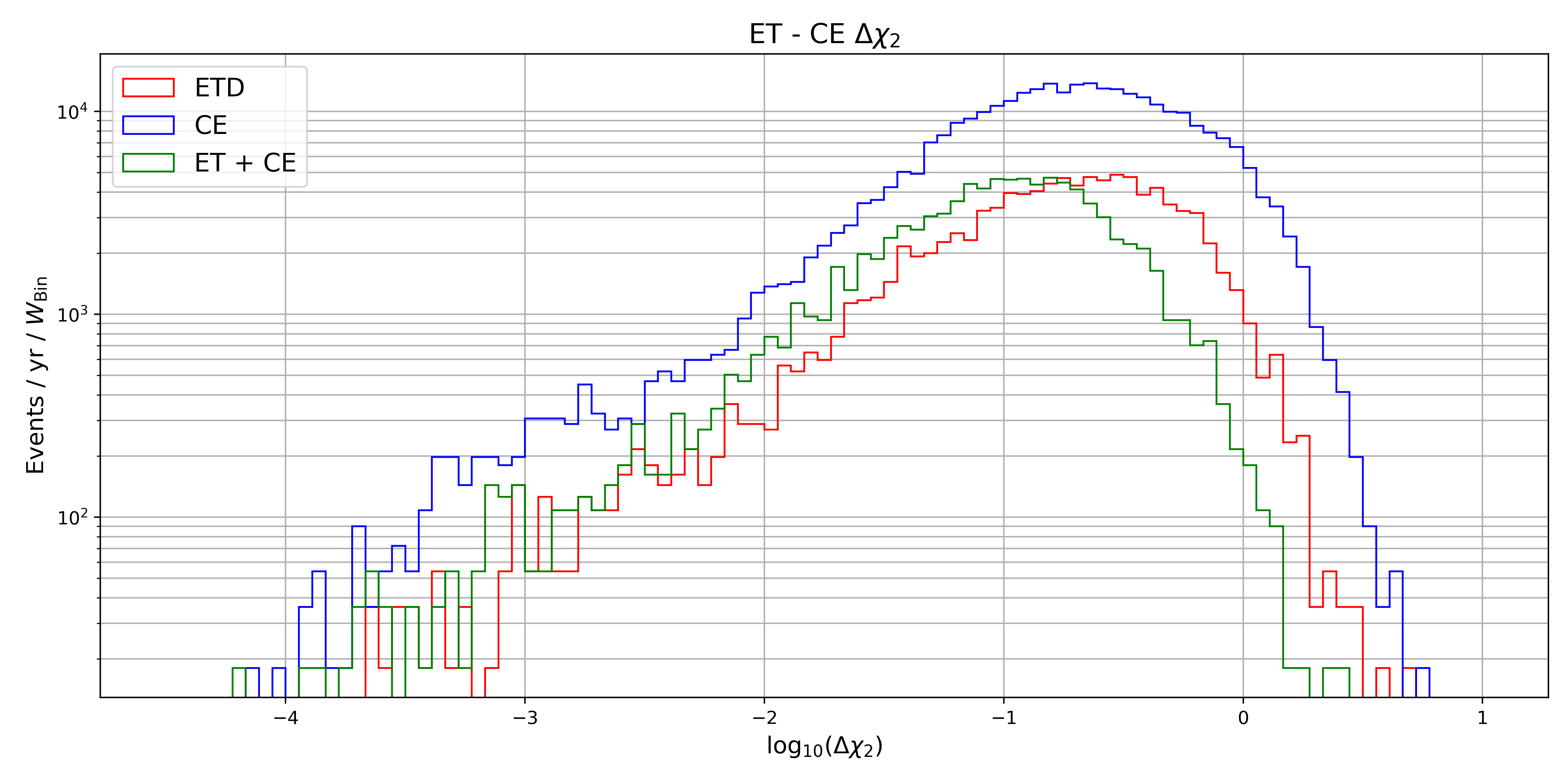}
	\caption{ \it{	Projected relative error on the second spin.}}
	\label{fig:spin_diff}
\end{figure}

\section{\sc Parameter estimation }
\label{sec:ex_par}

As shown in the previous section, the CE detector allows for a larger number of detected sources, compared to ET. This can be also seen in Fig.~\ref{fig:sensitivities}, where we plot the distribution of the  events detected in one year, for ET and CE separately and for them working as a network. We can note again that CE has a slightly better ability to detect BBH sources at higher redshift;  for ET it is expected that a large number of detected sources will be around $z\sim 1$, while for CE the maximum is at slightly larger redshift.

In Fig.~\ref{fig:SNR_mass_z} we also show the distribution of SNR for the detected binaries. One can see that a large number of events will have an SNR above $50$. This is consistent with \cite{Grimm:2020ivq}, confirming that ET and CE will be  more sensitive than currently operating ground-based and future space-based detectors when it comes to detecting SOBHBs. As an example, in Fig.~\ref{fig:SNR_mass_z_2} we also show the distribution of source-frame mass and redshift for our simulated binaries (for one realization of the universe), together with the SNR (color coded),  for the network ET+CE. One can see that the distribution in mass qualitatively follows the shape of the power-law+peak mass function, while the distribution of events in redshift has a maximum at $z \sim 2$. As expected, the SNR is larger for closer and more massive objects.

\begin{figure}[t!]
	\hspace{1cm} 
	\hspace{-1.5cm}\includegraphics[width=1.05\textwidth]{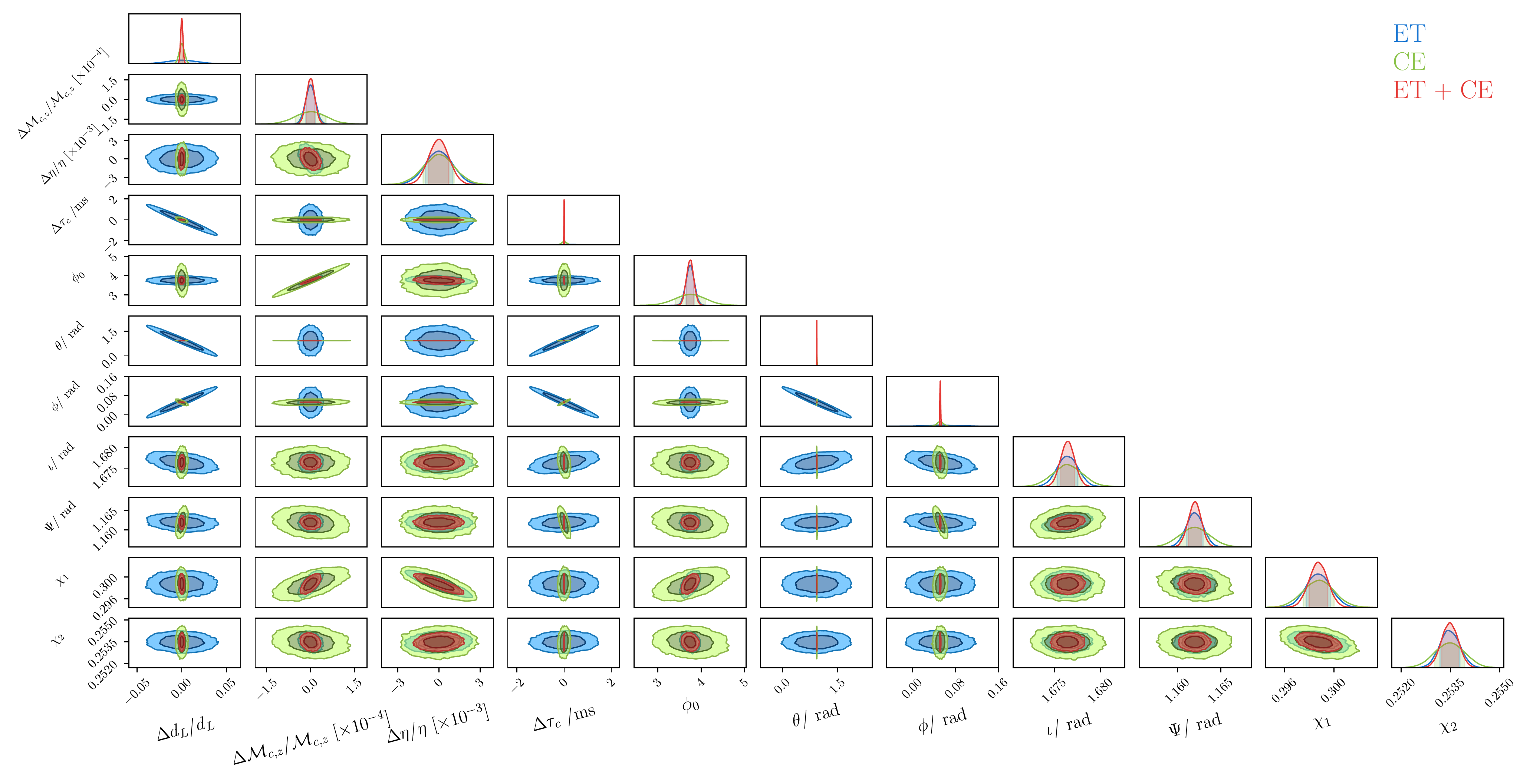}
	\caption{\label{fig:corner}{\it Fisher-matrix posteriors for ET, CE and the network ET+CE, for one of the loudest system (i.e., ${\rm SNR}_{\rm ET}$ $ = 607$, ${\rm SNR}_{\rm CE}$ $ = 596$) in one of our realizations of the universe, i.e. one with$d_L \simeq 0.57$ Gpc, $m_1\simeq 38.28 M_\odot$, $m_2\simeq 28.18 M_\odot$,  $\tau_c \simeq 0.06 $ yrs, $\phi_0\simeq 3.74$, $\theta\simeq 0.92$, $\phi\simeq 0.05$, $\iota\simeq 1.67$, 
			$\psi \simeq 1.16$, $S_1\simeq 0.49 $, $S_2\simeq 0.40$, $\cos(\theta_1) \simeq 0.60 $, $\cos(\theta_2) \simeq 0.61$. For practical reasons, the central value of $\tau_c$ here is set to zero. The credible levels in the plot represent the $1$ and $2\sigma$ regions for the astrophysical parameters.}}
\end{figure}

We will now compare the  parameter estimation capabilities of ET and CE, which are an important metric to assess the relative performance of different detector designs.
 As an example, in Fig.~\ref{fig:corner} we show the Fisher matrix posteriors for one of the loudest BH binary system (with ${\rm SNR}_{\rm ET}$ $ = 607$, ${\rm SNR}_{\rm CE}$ $ = 596$) in a realization of our population, i.e. one with
$d_L \simeq 0.57$ Gpc, $m_1\simeq 38.28 M_\odot$, $m_2\simeq 28.18 M_\odot$,  $\tau_c \simeq 0.06 $ yrs, $\phi_0\simeq 3.74$, $\theta\simeq 0.92$, $\phi\simeq 0.05$, $\iota\simeq 1.67$, 
$\psi \simeq 1.16$, $S_1\simeq 0.49 $, $S_2\simeq 0.40$, $\cos(\theta_1) \simeq 0.60 $, $\cos(\theta_2) \simeq 0.61$, firstly comparing ET with CE and then combining the two detectors. As expected, CE, having two detectors at different locations, allows for estimating the extrinsic parameters (e.g. $d_L$ and sky position) more precisely. This is of course due to the possibility of ``triangulating'' the sources in the sky.
We show this also in Fig.~\ref{fig:extrinsic}, where we plot the projected error on the sky localization, luminosity distance and inclination for the whole SOBHB population. We can see that the two L-shaped CE detectors allow for a better determination of the sky position compared to the single ET triangular detector. However, in both cases we can reach a resolution of order $5-10\, {\rm{deg}}^2$ for a quite large number of detected events. The combination of the two allows to reach accurate angular resolution up to percent level.
On the other hand, the difference on the luminosity distance error is not so large, even if CE allows a good estimation for a larger number of events. A similar behavior holds also for the inclination angle.

To be more quantitative, in Table \ref{tab: number_om} we have reported the number of sources that can be localized in the sky within 10, 1 and 0.1 square degrees, for the two detectors and for the network at low (i.e., $z<2$) and high (i.e., $z>2$) redshift, for one year of observation. This information is crucial to understand how many detections can be useful for cosmography and for cross-correlation with galaxy surveys~\cite{Sathyaprakash:2009xt}. 
For comparison, we note that a 10 $\rm deg^2$ error box is the field of view of SKA~\cite{ska} or Vera Rubin Observatory (LSST)~\cite{verarubin}.
One can see that the difference in localization power between ET or CE alone compared to their combination is quite significant both at low and high redshift.

\begin{table}[h!]
\centering
\begin{tabular}{ cc }  
$z<2$ & $z>2$\\  
\begin{tabular}{ |c|c|c|c|c| } 
\hline
{}       & ET & CE & ET + CE \\
\hline
\hline
$\# {\rm events}$  &  4769 & 13302 & 4372  \\
\hline
$\Delta \Omega <10$  &  6.5\% & 16\% & 99.8\%  \\
\hline
$\Delta \Omega <1 $ &  0.44\% & 1.62\% & 59.7\% \\
\hline
$\Delta \Omega <0.1 $ & 0.\% & 0.05\%  & 5.9\% \\
\hline
\end{tabular} & 
\quad\quad 
\begin{tabular}{ |c|c|c|c|c| }
\hline
{}       & ET & CE & ET + CE  \\
\hline
\hline
$\# {\rm events}$  &  872 & 4388 & 805  \\
\hline
$\Delta \Omega <10$   &  4.47\% & 5.4\% & 98\% \\
\hline
$\Delta \Omega <1$  & 0\% & 0.2\% & 26.5\% \\
\hline
$\Delta \Omega <0.1$  & 0\% & 0\% & 0\% \\
\hline
\end{tabular}
\end{tabular}
\caption{{\it Percentage of events that can be detected with relative errors within 10, 1 and 0.1 square degrees on the sky localization angle. The left table refers to redshifts below $z = 2$, while right one refers to redshifts above $z = 2$}
\label{tab: number_om}}
\end{table}

\begin{figure}[t!]
	\centering
		\includegraphics[width=.9\textwidth]{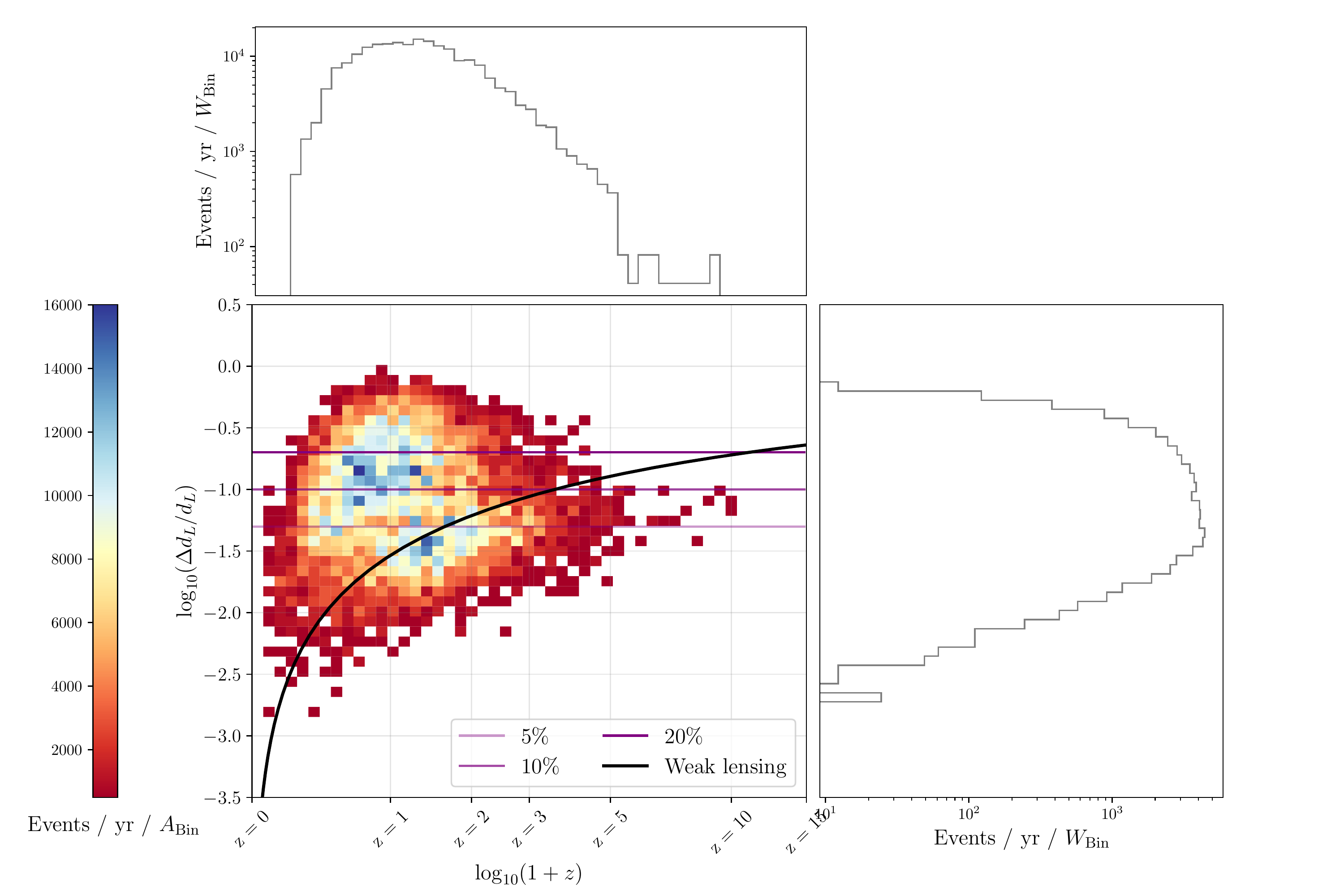}
	\caption{ \it{	Projected relative error on the luminosity distance coming from the weak lensing compared with the  statistical error on the distance determination  for the network ET+CE.}}
	\label{fig:weak_lens}
\end{figure}

We have also estimated the percentage of events that can be detected with relative  errors on the luminosity distance of $20\%$, $10\%$ and $5\%$ for redshift below and above $z = 2$, including the systematic error due to weak lensing.
From Table \ref{tab: number_dL}, one can see that at low redshift a reasonable number of events are detectable with a $20\%$ accuracy in the three  configurations (ET, CE and ET+CE). However, as more stringent requirements on the determination of the luminosity distance are considered, one can clearly see the benefit of having a network of detectors. Note that at high redshift, however, no events are detectable with accuracy below $5\%$ and just a few with $10\%$ accuracy. An explanation for this limitation can be obtained from the plot in Fig.~\ref{fig:weak_lens}, where we have compared for the network ET+CE the statistical error on the distance with the weak lensing contribution, which becomes dominant at high $z$.

\begin{table}[htb!]
	\centering
	\begin{tabular}{ cc }  
		$z<2$ & $z>2$\\  
		\begin{tabular}{ |c|c|c|c |c } 
			\hline
			{}       & ET & CE & ET + CE  \\
			\hline
			\hline
			$\# {\rm events}$  &  4769 & 13302 & 4372  \\
			\hline
			$\Delta d_L < 20 \% $  &  16\% & 21\% & 75\% \\
			\hline
			$\Delta d_L < 10 \% $ &  4.4\% & 7\% & 45\% \\
			\hline
			$\Delta d_L < 5 \% $ & 0.77\% & 1.3\% & 13\% \\
			\hline
		\end{tabular} &  
		\quad
		\begin{tabular}{ |c|c|c|c|c } 
			\hline
			{}       & ET & CE & ET + CE  \\
			\hline
			\hline
			$\# {\rm events}$  &  872 & 4388 & 805  \\
			\hline
			$\Delta d_L < 20 \% $  &  11\% & 11.3\% & 76\% \\
			\hline
			$\Delta d_L < 10 \% $  & 0\% & 0.3\% & 8.3\% \\
			\hline
			$\Delta d_L < 5 \% $  & 0\% & 0\% & 0\% \\
			\hline
		\end{tabular} \\
	\end{tabular}
	\caption{{\it Percentage of events that can be detected with relative errors on the luminosity distance of $20\%$, $10\%$ and $5\%$. The left table refers to redshifts below $z = 2$, while the right one refers to redshifts above $z = 2$.}
		\label{tab: number_dL} 
	}
\end{table}

The parameter estimation results for the intrinsic parameters (i.e., detector-frame chirp mass, symmetric mass ratio and  spins) are shown 
in Fig.~\ref{fig:intrinsic}. As can be seen, the impact of the detector choice is 
less significant than for the extrinsic parameters. 
In more detail, with both detectors considered in this work, the chirp mass is always estimated to sub-percent error or better, with a slightly better error given by CE. 
The symmetric mass ratio and the spins are estimated to within 1\% or better for most detected events, and   for the symmetric mass ratio  a large fraction of sources are expected to be measured even with
sub-percent error. Finally in Fig.~\ref{fig:spin_diff} we plot the projected relative error on the difference between the two spins.

In our analysis we have not neglected correlations between the parameters; in fact, in Fig.~\ref{fig:corner} it can be noticed that there are some correlations among some parameters; for instance it is manifest that there is a degeneracy $\theta - \phi$. Two important remarks about the correlation are the followings: i) to produce Fig.~\ref{fig:corner},  we have chosen one of the loudest events (i.e., SNR $\simeq$ 600) which of course translate in tighter errors on several parameters; ii) in our analysis we have included higher order modes in the waveform which help in reducing some degeneracies.

\section{\sc Conclusions}
Among the most interesting proposals to improve the sensitivity of current ground-based GW interferometers  there are the ET and CE detectors. These interferometers will be characterized by longer arm-lengths and an order of magnitude better sensitivity compared to the LIGO/Virgo/KAGRA detectors. They 
are currently planned to 
 have a triangular and L-shape geometry, respectively. With these characteristics,  they will offer the possibility of exploring our universe up to high redshift and to detect a large number of GW sources, shedding light on the population properties of  compact objects through the dark age of our universe. In this paper, we have studied the detectability and  parameter estimation capabilities  of ET and CE, considering them both independently or in a network. We have developed synthetic catalogues for one of the primary targets of ET and CE -- SOBHBs -- accounting for the latest LIGO/Virgo constraints on the population properties. We have used a redshift-dependent merger rate and a power-law plus peak mass function.
 
We have first estimated the number of detectable events with the two detectors and with the network, and we have found that both options offer the chance to detect a large number of  SOBHBs up to redshift $z \sim 5-6 $. When the two instruments are considered in a network, almost all the SOBHBs up to $z\sim 1.5$ can be individually detected. ET and CE will also detect these systems with a high SNR, which will help in determining the source parameters.
  
Using a Fisher matrix analysis, we have then estimated the error on the intrinsic and extrinsic source parameters. As expected, the two \rm{L}-shaped CE interferometers have better sky localization capabilities, while the errors on the luminosity distance $\Delta d_L$ and the inclination angle $\Delta \iota$ are comparable. We have also quantified the number of detectable events with given sky resolution ($\Delta \Omega < 10, 1, 0.1$ square degrees) 
at low and high redshift ($z<2$ and $z>2$).
We have found that the network ET+CE will have a good angular resolution (better than 10 square degrees) up to $z\lesssim$ 3-4, which may allow for cross-correlating SOBHB events with galaxy surveys such as SKA or the Vera Rubin Observatory. The number of detectable events
with good angular resolution, however, 
drastically decreases at higher $z$.
A similar analysis 
for the luminosity distance 
shows that the number of events with fixed error (better than $20\%$, $10\%$ and $5\%$)
 is always larger for the ET detector.
 Moreover, the network ET+CE 
allows for a very accurate (10\% or better) measurement of the luminosity distance for almost half of the detected events.
 At high redshift, especially at $z\gtrsim 3$, the error on the distance increases due to weak lensing.
As for the intrinsic source parameters, the error on the chirp mass is slightly better for CE. As for the
symmetric mass ratio and the spins, errors are 
projected to be comparable for ET and CE, with CE allowing an accurate estimation for a larger number of events. When a network of the two detectors is considered, the errors further improve down to sub-percent levels.\\
One limitation of the current work consists in the assumption of non-overlapping signals. The event rate that we find for SOBHBs implies a detection every few hours, which would imply the possibility to have many of them overlapping. On the other hand we did not include binaries of NSs, which are also a prominent source for ET and CE.  We leave the inclusion of NS sources and overlapping signals for future works.

\acknowledgments
We thank P. Pani, A. Riotto and J. Zhang for useful feedback on the draft. We thank M. Mancarella and F. Iacovelli for useful comments and discussions, for comparing codes and for helping to spot a bug in a previous version of the code. E.B. acknowledges financial support provided under the European Union's H2020 ERC Consolidator Grant ``GRavity from Astrophysical to Microscopic Scales'' grant agreement no. GRAMS-815673. This work was supported by the EU Horizon 2020 Research and Innovation Programme under the Marie Sklodowska-Curie Grant Agreement No. 101007855. The work of M.P. was supported by STFC grants ST/P000762/1 and ST/T000791/1. M.P. acknowledges support by the European Union’s Horizon 2020 Research Council grant 724659 MassiveCosmo ERC- 2016-COG. A.R. acknowledges funding from MIUR through the ``Dipartimenti di eccellenza'' project Science of the Universe.

\newpage
\appendix
    
    \section{\sc Astrophysical model}
    \label{app:astr_mod}

The Power-Law+Peak mass distribution that we use is described by
\begin{eqnarray}
    \pi(m_1 |\lambda_\text{peak}, \alpha, \mmin, \delta_m, \mmax, \mu_m, \sigma_m) &=&
    \bigg[
    (1-\lambda_\text{peak})\mathfrak{P}(m_1|-\alpha, \mmax) +
    \lambda_\text{peak} G(m_1|\mu_m,\sigma_m)
    \bigg]\nonumber\\
   &&\times S(m_1|\mmin, \delta_m) ,
\end{eqnarray}
where $\mathfrak{P}(m_1|-\alpha,\mmax)$ is a normalized power-law distribution with spectral index $-\alpha$ and high-mass cut-off $\mmax$, and $G(m_1|\mu_m, \sigma_m)$ is a normalized Gaussian distribution with mean $\mu_m$ and width $\sigma_m$.
$S(m_1, \mmin, \delta_m)$ is a smoothing function, which rises from 0 to 1 over the interval $(m_{min}, m_{min}+\delta_m)$:
\begin{equation}
\label{eq:smoothing}
S(m \mid \mmin, \delta_m) = \begin{cases}
    0 & \left(m< \mmin\right) \\
    \left[f(m - \mmin, \delta_m) + 1\right]^{-1} & \left(\mmin \leq m < \mmin+\delta_m\right) \\
    1 & \left(m\geq \mmin + \delta_m\right)
\end{cases}
\end{equation}
with
\begin{equation}
    f(m', \delta_m) = \exp \left(\frac{\delta_m}{m'} + \frac{\delta_m }{m' - \delta_m}\right).
\end{equation}
The conditional symmetric mass ratio distribution in this model also includes the smoothing term:
\begin{align}
\label{eq:pq_smoothing}
\pi(q \mid \beta, m_1, \mmin, \delta_m) \propto q^{\beta_q} S(q m_1 \mid \mmin, \delta_m).
\end{align}
We have chosen the same parameters as  in the LIGO/Virgo/KAGRA paper where this parametrization was introduced~\cite{LIGOScientific:2020kqk}. \\
The spin of each BH component in a binary is assumed to be independently drawn from a common underlying distribution, and  the dimensionless spin magnitude is described using a beta distribution:
\begin{align}
    \pi(\chi_{1,2} | \alpha_\chi, \beta_\chi) = \text{Beta}(\alpha_\chi, \beta_\chi) ,
\end{align}
where $\alpha_\chi$ and $\beta_\chi$ are shape parameters that determine the distribution's mean and variance.\\
In Table \ref{tab:Astro_par} we report the values of the parameters

\begin{table}
\centering
\begin{tabular}{|c|c|}
\hline
Parameters & Value \\
\hline
$m_{min} $ & $ 2.5 M_{\odot}$\\
\hline
$m_{max} $ & $ 100 M_{\odot}$\\
\hline
$\delta_{min} $ & $ 7.8^{+1.9}_{-4.0} M_{\odot}$\\
\hline
$\lambda_{peak}$ & $ 0.039^{+0.058}_{-0.026} $\\
\hline
$\alpha$ & $ 3.4^{+0.58}_{-0.49} $\\
\hline
$\beta_q$ & $ 1.1^{+1.8}_{-1.3}$\\
\hline
$\mu_m$ & $ 34^{+2.3}_{-3.8} $\\
\hline
$\sigma_m$ & $ 5.09^{+4.28}_{-4.34}$\\\hline
\end{tabular}
\caption{{\it Table of the parameters used in the distribution functions, taken from~\cite{LIGOScientific:2021psn}}}.
\label{tab:Astro_par}
\end{table}

\bibliographystyle{JHEP}
\bibliography{references.bib}

\end{document}